\documentclass[aps,amsmath,amssymb,twocolumn]{revtex4}
\usepackage{blindtext}

\usepackage{graphicx}
\usepackage{dcolumn}
\usepackage{bm}
\usepackage{amsmath}
\usepackage{physics}
\usepackage{braket}
\usepackage{xcolor}
\usepackage{amsfonts}

\begin{document}

\preprint{APS/123-QED}

\title{Speeding up Lindblad dynamics via time-rescaling engineering}

\author{Bertúlio de Lima Bernardo}%
 \email{bertulio.fisica@gmail.com}
\affiliation{%
Departamento de F\'{i}sica, Universidade Federal da Para\'{i}ba, 58051-900 Jo\~{a}o Pessoa, PB, Brazil
}%

\date{\today}

\begin{abstract}
We introduce a universal method for accelerating Lindblad dynamics that preserves the original trajectory through Hilbert space. The technique provides exact fast processes analytically, which are Markovian and do not require manipulation of the environment properties, by time-rescaling a reference dynamics. In particular, the engineered control protocols are based only on local interactions, and no additional control fields are required compared to the reference protocol. We demonstrate the scheme with two examples: a driven two-level system in an amplitude damping channel and the dissipative transverse field Ising model. We also show that, by starting with a reference process which is the fastest connecting two states under a certain constraint, the method provides other optimal processes satisfying modified constraints. Our approach can help advance techniques for quantum control and computation towards more complex noisy systems.       
\end{abstract}

                    
\maketitle


\section{Introduction}

The main challenges to improve the performance of current noisy intermediate-scale quantum (NISQ) devices are to increase the number of qubits involved and devise operations that are completed within the coherence time \cite{preskill}. Thus, tremendous efforts have been recently devoted to developing quantum control techniques intended to precisely manipulate more quantum units in shorter time intervals \cite{ippoliti}. Shortcuts to adiabaticity (STA) address this issue by providing fast quantum dynamics that have the same outcome as slow adiabatic processes \cite{odelin,torrengui}. Examples of STA methods proposed so far include counterdiabatic driving (CD) \cite{demi,demi2,berry}, invariant-based inverse engineering \cite{chen}, fast-forward approach \cite{masuda,masuda2}, and, more recently, time-rescaling \cite{bernardo}. As originally proposed, these techniques have been shown to be efficient in speeding up closed quantum dynamics \cite{bason, zhang,kumar,roh,diao,roy,andrade,ferreira}.  

During the last ten years, progress has been made in extending STA to open quantum systems \cite{vacanti,santos,alipour,dann,funo}, as demonstrated in a recent experiment using a superconducting circuit architecture \cite{yin}. Inverse engineering strategies have been used in some of these extensions. For instance, in Ref.~\cite{alipour} the authors prescribe a target trajectory for the evolution of the system, from which the corresponding master equation is constructed. In this case, speedups of strokes in an open two-level system are studied, where the obtained dynamics are either non-Markovian with time-dependent Lindblad operators or described in terms of a non-Hermitian Hamiltonian. In \cite{dann}, a reverse-engineering idea is used to speed up the relaxation of open quantum systems towards the equilibrium state, in which the thermalization of a particle in a harmonic well is investigated. Other efforts to extend the applicability of STA to open quantum systems have predominantly focused on the CD method. Based on a specific formulation of adiabaticity for systems undergoing open evolution, CD was first generalized in \cite{vacanti}. In a subsequent work \cite{santos}, an extension that provides phase freedom in the Liouvillian superoperator was introduced. 

Similarly to the applications of STA to closed quantum systems, the counterparts to open systems have been shown to be efficient at manipulating simple systems. However, attempts to use these techniques in both closed and open many-body systems have suffered from serious problems. First, the implementation of STA in this case often requires non-local infinite-range interactions, therefore precluding experimental realizations. Second, the design of CD protocols for closed and open systems requires knowledge of the instantaneous spectrum of the reference Hamiltonian and Liouvillian superoperator, respectively, which evidently becomes impractical for many-body dynamics. To circumvent these limitations, a variational approach was introduced for closed quantum systems that allows an approximate construction of the CD Hamiltonian \cite{sels}. This approximate method was recently extended to open quantum systems \cite{passarelli}. 

Here, we report a universal method to speed up a reference Lindblad dynamics through the same trajectory in Hilbert space, which is a generalization of the time-rescaling method \cite{bernardo} to open quantum systems. The scheme analytically provides exact solutions for the fast dynamics, which are Markovian and are generated by time-independent environments. Remarkably, the computation of the fast processes does not require knowledge about the eigenvalues of the Liouvillian, the experimental implementation involves only local interactions, and no extra fields are required with respect to the reference protocol. We also show that the time evolution of the fast processes relates to the reference by a simple reparametrization of time, and provide a discussion on the quantum speed limit associated with the resulting dynamics. Our framework is demonstrated for two cases: a two-level system in an amplitude damping channel and the dissipative transverse field Ising model on two sites.  

The article is organized as follows. In Sec.~\ref{theory}, we introduce the time-rescaling method for accelerating Lindblad dynamics and demonstrate how the reference process is related to the resulting accelerated dynamics. In Sec.~\ref{results1}, we apply the method to derive fast evolutions for a driven two-level system under amplitude damping and for the dissipative transverse-field Ising model. Section~\ref{results2} revisits these examples, this time deriving accelerated dynamics that do not require time-dependent control of the environment. In Sec.~\ref{qslimit}, we investigate how the quantum speed limit properties of the reference process  impact those of the accelerated processes. Finally, our conclusions are presented in Sec.~\ref{conclusions}.

\section{Theory and methods}
\label{theory}

In the Markovian regime, the dynamics of an open quantum system is described by the Lindblad equation,
\begin{eqnarray}
\frac{d \rho}{d t} = \mathcal{L} \rho,
\end{eqnarray}
where
\begin{eqnarray}
\mathcal{L} \rho = -\frac{i}{\hbar} [H,\rho] + \sum_{k} \gamma_{k} \left( L_{k} \rho L^{\dagger}_{k} - \frac{1}{2} \{L^{\dagger}_{k} L_{k}, \rho \}\right). 
\end{eqnarray}
Here, $\rho$ is the density operator, $\mathcal{L}$ is the Liouvillian superoperator, and $H$ is the system Hamiltonian, which generates the unitary part of the dynamics. The Lindblad operators $L_{k}$ characterize nonunitary transformations
such as relaxation and decoherence that take place at rates $\gamma_{k} \geq 0$ for all time $t$ \cite{breuer,gorini,lindblad}. For clarity, we omitted the time dependence of $\rho(t)$, $\mathcal{L}(t)$, $H(t)$, $L_k(t)$, and $\gamma_k(t)$ in the expressions above. The Lindblad equation represents the general description of the time evolution of the state of a system under the action of a completely positive trace-preserving (CPTP) map \cite{campaioli}. Usually, it is derived under the assumptions of weak system-environment coupling and the absence of memory effects (Born-Markov approximation) \cite{lidar}.  

When solving the Lindblad equation, it is often convenient to rewrite it in the vectorized form \cite{havel}:  
\begin{equation}
\label{veclindblad}
\frac{d \ket{\rho}}{d t} = \mathcal{L} \ket{\rho},
\end{equation}
where $\ket{\rho}$ is a $d^2$-entry column vector obtained by stacking the columns of the $d \times d$ density operator $\rho$, with $d$ being the dimension of the Hilbert space, and the Liouvillian superoperator becomes
\begin{eqnarray}
\mathcal{L} = &-&\frac{i}{\hbar}(\mathbb{I} \otimes H - H^T \otimes \mathbb{I}) \nonumber \\
&+& \sum_{k} \gamma_k \left[L^{*}_{k} \otimes L_{k} - \frac{1}{2} \mathbb{I} \otimes L^{\dagger}_{k}L_{k} - \frac{1}{2} (L^{\dagger}_{k}  L_{k})^{T} \otimes \mathbb{I}  \right]. \nonumber \\
\end{eqnarray}
The superscripts $*$, $T$, and $\dagger$ represent the complex conjugate, transpose, and adjoint operators, respectively, and $\mathbb{I}$ is the identity operator. The general solution of Eq.~(\ref{veclindblad}) is given by $\ket{\rho(t_f)} = \Lambda(t_f,0) \ket{\rho(0)}$, with
\begin{equation}
\label{propagator}
\Lambda(t_f,0) = \mathcal{T}\exp \left\{ \int_{0}^{t_{f}} \mathcal{L}(t') dt' \right\}
\end{equation}
being the propagator of the dynamics that starts at time $t = 0$ and finishes at $t = t_f$ \cite{campaioli,lidar}. Here, $\mathcal{T}$ is the time-ordering operator, similar to the Dyson series for time-dependent Hamiltonian dynamics \cite{sakurai}. We call the dynamics generated by $\mathcal{L}(t)$ the {\it reference process}, and point out that it takes a time $\Delta t_ {ref} = t_f$ to occur.  

The main goal of this work is to design new processes that are capable of driving the system from the same initial state $\ket{\rho(0)}$ to the same final state $\ket{\rho(t_f)}$ in a shorter time interval, $\Delta t < \Delta t_ {ref}$. For that, we consider the change $t'=f(\tau)$ in the time variable of Eq.~(\ref{propagator}), where we call $f(\tau)$ the {\it time-rescaling function}. This change transforms the propagator as follows
\begin{align}
\label{TRpropagator}
\tilde{\Lambda}[f^{-1}(t_{f}),0] &= \mathcal{T} \exp \left\{ \int_{f^{-1}(0)}^{f^{-1}(t_{f})} \mathcal{L}[f(\tau)]\dot{f}(\tau) d\tau \right\} \nonumber \\
&= \mathcal{T} \exp \left\{  \int_{f^{-1}(0)}^{f^{-1}(t_{f})} \tilde{\mathcal{L}}(\tau) d\tau \right\}.
\end{align}
Later we will require that $f^{-1}(0) = 0$ such that the time duration of the new process becomes $\Delta t = f^{-1}(t_{f})$. We observe that the last expression in Eq.~(\ref{TRpropagator}) has the same form as Eq.~(\ref{propagator}). This means that we obtained a modified propagator with the same effect as that of the reference process, which starts at time $t = 0$ and ends at $t = f^{-1}(t_{f})$. The Liouvillian that generates the new process relates to that of the reference process simply by  
\begin{align}
\label{liouvillians}
\tilde{\mathcal{L}}(t) = \mathcal{L}[f(t)]\dot{f}(t).
\end{align}
The new process will be called the {\it time-rescaled} (TR) {\it process}, where $f^{-1}(t)$ and $\dot{f}(t)$ are the inverse and the first derivative of $f(t)$, respectively. 

It is important to observe that the propagator of Eq.~(\ref{TRpropagator}) was obtained from that of Eq.~(\ref{propagator}) simply by a change of variable, i.e., both have exactly the same effect. That is, the transformation caused by $\tilde{\Lambda}[f^{-1}(t_{f}),0]$ in an arbitrary initial state $\ket{\rho(0)}$, which takes a time $\Delta t$, 
is the same transformation caused by $\Lambda(t_f,0)$, which takes a time $\Delta t_{ref}$. In mathematical terms, we have $\ket{\rho(t_f)} = \tilde{\Lambda}[f^{-1}(t_{f}),0] \ket{\rho(0)}= \Lambda(t_f,0) \ket{\rho(0)}$. Naturally, if we are interested in a TR process that is faster than the reference, we still need to guarantee that $\Delta t < \Delta t_{ref}$.

A key challenge in controlling open quantum dynamics is to design fast processes in which the system experiences specific external conditions at the beginning and end of the evolution. This is crucial, for instance, in the execution of noisy quantum computations with prescribed boundary conditions \cite{knill,brav}, or in implementing nonunitary strokes in quantum heat engines \cite{myers}. To satisfy this requirement, the Liouvillian superoperators of the reference and TR processes must coincide at the initial and final times. This ensures that the system is subjected to the same Hamiltonians parameters and environmental conditions at the start and end of both processes. For that, $f(t)$ must satisfy the following properties: (i) $f^{-1}(0) = 0$, for the initial times of the reference and the TR processes to be equal; (ii) $f^{-1}(t_{f})<t_{f}$, for the TR process to be faster; (iii) $\tilde{\mathcal{L}}(0) = \mathcal{L}(0)$, to make the initial Liouvillians equivalent; and (iv) $\tilde{\mathcal{L}}[f^{-1}(t_{f})] = \mathcal{L}(t_{f})$, to make the final Liouvillians equivalent.

Similar to the case of closed quantum dynamics \cite{bernardo}, we have that the function
\begin{equation}
\label{TRF}
f(t) = a t - \frac{(a-1)}{2 \pi a} t_{f} \sin \left( \frac{2 \pi a}{t_{f}} t \right)
\end{equation}
meets all the requirements above. Indeed, for $a>1$ we find that the evolution of the TR process is equivalent to the reference one, but occurs $a$ times faster, say $\Delta t = \Delta t_{ref} / a$. For this reason, we call $a$ the {\it time contraction parameter}. Next, we show that the trajectory followed by the state of the system in Hilbert space is the same for both the reference and the TR processes.  

\begin{figure}[htb]
\begin{center}
\includegraphics[width=0.44\textwidth]{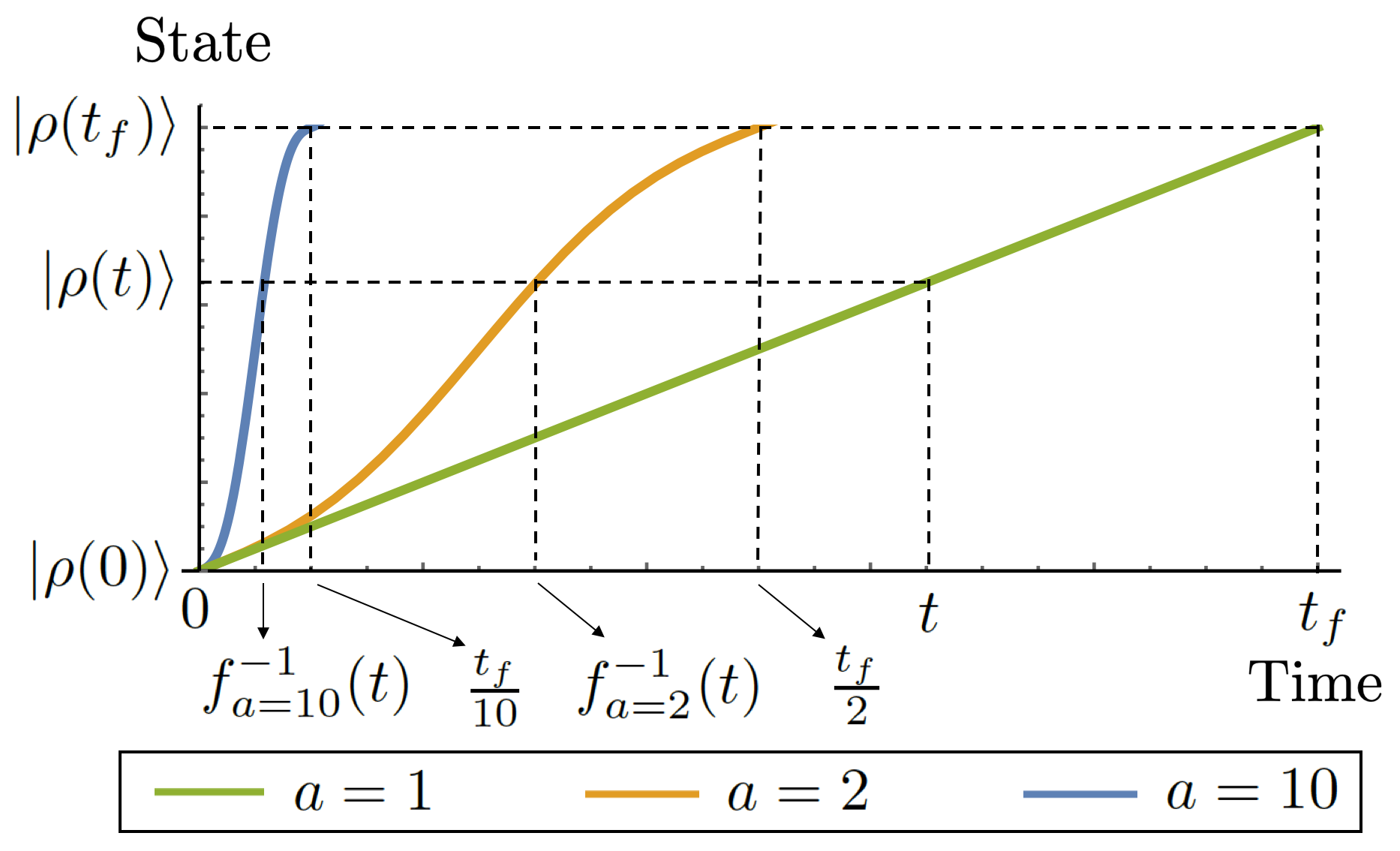}
\caption{Time evolution of a reference ($a=1$) Lindblad dynamics and the corresponding TR dynamics for $a=2$ and $a=10$. In all cases, the trajectories from $\ket{\rho(0)}$ to $\ket{\rho(t_f)}$ in Hilbert space are the same. The time duration of the TR processes are $a$ times faster than the reference one.}
\label{TRroutes}
\end{center}
\end{figure}

Let us introduce a parameter $\eta \in [0,1]$ which characterizes the stage of the trajectory followed by the density operator in the reference process according to $\ket{\rho(\eta t_f)} = \Lambda(\eta t_f,0) \ket{\rho(0)}$, where 
\begin{equation}
\label{stagepropagator}
\Lambda(\eta t_f,0) = \mathcal{T}\exp \left\{ \int_{0}^{\eta t_{f}} \mathcal{L}(t') dt' \right\}.
\end{equation}
Here, $\ket{\rho(\eta t_f)}$ represents all states of the trajectory generated by the Liouvillian $\mathcal{L}(t)$, from $\ket{\rho(0)}$ (when $\eta = 0$) to $\ket{\rho(t_f)}$ (when $\eta = 1$). Now, if we apply the same change of variable as before, say  
$t'=f(\tau)$, we get
\begin{align}
\label{stageTRpropagator}
\tilde{\Lambda}[f^{-1}(\eta t_{f}),0] 
= \mathcal{T} \exp \left\{  \int_{0}^{f^{-1}(\eta t_{f})} \tilde{\mathcal{L}}(\tau) d\tau \right\}.
\end{align}
Since $\tilde{\Lambda}[f^{-1}(\eta t_{f}),0]$ was obtained from $\Lambda(\eta t_f,0)$ by a change of variable, both have the same effect if applied to an arbitrary initial state. Namely, $\ket{\rho(\eta t_f)} = \Lambda(\eta t_f,0) \ket{\rho(0)} = \tilde{\Lambda}[f^{-1}(\eta t_{f}),0] \ket{\rho(0)}$. In other words, all states of the reference dynamics $\ket{\rho(\eta t_f)}$, which are reached in a time $\eta t_f$, are also obtained in the TR process in a shorter time $f^{-1}(\eta t_f)$. Yet, as there is a one-to-one correspondence between the propagators $\Lambda(\eta t_f,0)$ and $\tilde{\Lambda}[f^{-1}(\eta t_{f}),0]$ characterized by $\eta$, we learn that the trajectories produced by the reference and the TR processes are exactly the same. In Fig.~\ref{TRroutes} we sketch the relation between the trajectory states and the time at which they occur in the reference ($a=1$) and the TR processes for $a=2$ and $a=10$.   

Based on Fig.~\ref{TRroutes}, and adopting the simplification $t=\eta t_f$, we can express a relation between the reference and the TR dynamics as
\begin{equation}
\label{trroute}
\qquad  \qquad  \quad  \quad \ket{\tilde{\rho}[f^{-1}(t)]} =   \ket{\rho(t)}, \qquad  0 \leq t \leq t_f,
\end{equation}
where the superscript $\sim$ represents the TR dynamics. This remarkable result implies that the TR dynamics can be analytically obtained from the reference dynamics by a simple reparametrization of time. With this facility, starting from a known open quantum dynamics as a reference, one is able to directly compute the analogue fast dynamics without resorting to numerical calculations. This connection may be particularly useful for quantitative studies of many-body quantum dynamics \cite{daley,cemin,defenu}. 

\section{Results}
\label{results1}

\begin{figure*}[htb]
\begin{center}
\includegraphics[width=1.0\textwidth]{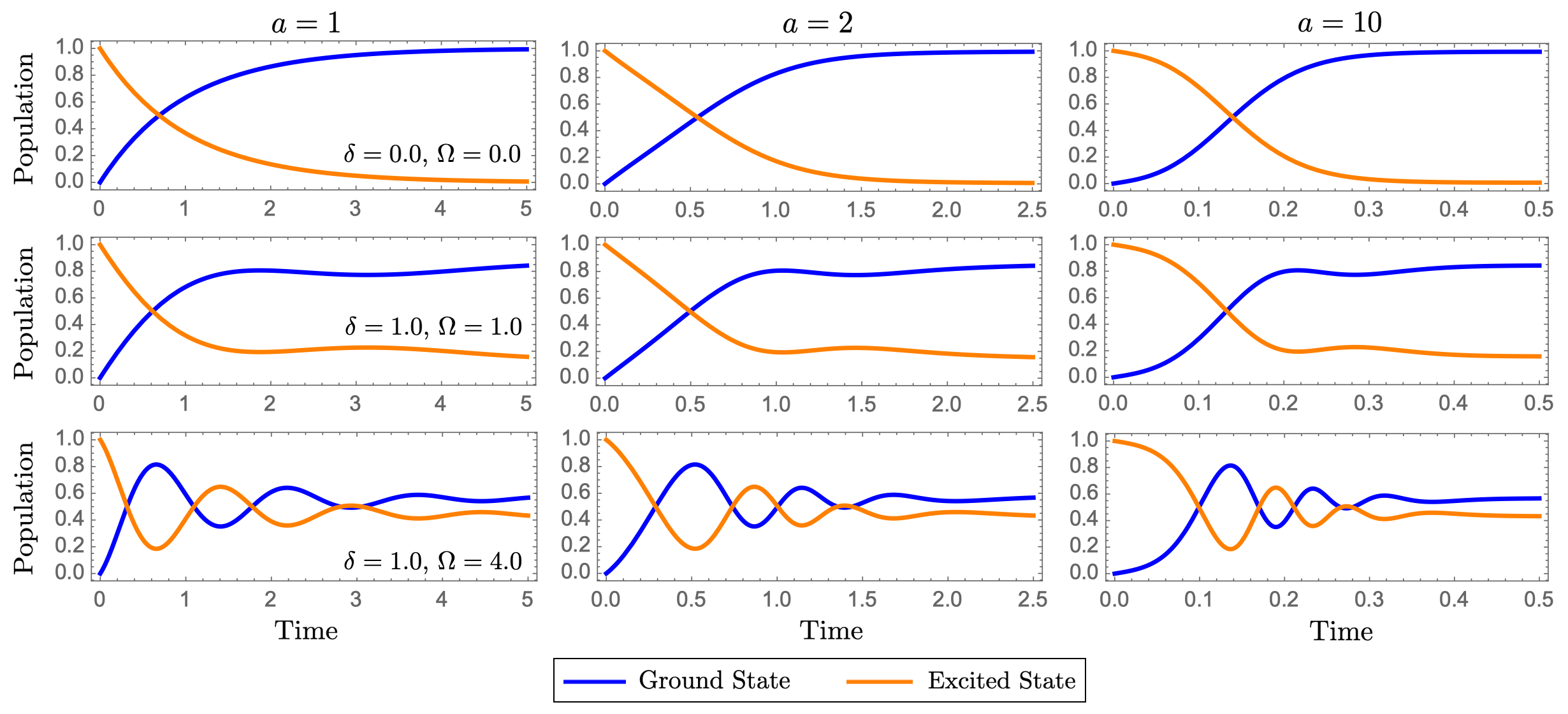}
\caption{Spontaneous emission of a driven two-level system. Ground (blue) and excited (orange) state population dynamics of a driven two-level system under the action of the amplitude damping channel. The detuning $\delta$ and the Rabi frequency $\Omega$ of the reference dynamics (first column) are set as constants, with the decay rate $\gamma =1$. The results of the engineered TR dynamics are shown in the second and third columns for the $a=2$ and $a=10$ cases, respectively. The system starts out in the excited state in all cases.}
\label{amplitudedamping}
\end{center}
\end{figure*}

We apply our method to two models of open system dynamics, and call attention that, in this section and the next, we assume $\hbar = 1$. First, we study the design of processes that accelerate the dynamics of a two-level system (TLS) under the action of the amplitude damping channel and driven by the Hamiltonian  
\begin{equation}
H = - \frac{\delta}{2} \sigma_z -\frac{\Omega}{2} \sigma_x, 
\end{equation}
where $\sigma_x$ and $\sigma_z$ are the $x$ and $z$ Pauli operators, $\delta$ is the detuning and $\Omega$ is the Rabi frequency. In this case, the action of the environment is described by the Lindblad operator $\sigma_{-}= \ketbra{0}{1}$, where $\sigma_{z} \ket{0}= \ket{0}$ and $\sigma_{z} \ket{1}= -\ket{1}$, and by the spontaneous emission rate $\gamma$. Here, we call $\ket{0}$ and $\ket{1}$ the ground and excited states, respectively. Fig.~\ref{amplitudedamping} shows the time evolution of the population of the states $\ket{0}$ and $\ket{1}$ in several different parameter regimes, with the initial state assumed to be $\ket{1}$. The reference ($a=1$) dynamics are displayed in the first column, where we set the time duration as $t_f = 5$ and $\gamma = 1$ in adimensional units. The detuning $\delta$ and the Rabi frequency $\Omega$ were varied in order to obtain qualitatively different dynamics. The results of the reference dynamics are in agreement with Refs.~\cite{kamakari,schli}. 

\begin{figure*}[htb]
\begin{center}
\includegraphics[width=1.0\textwidth]{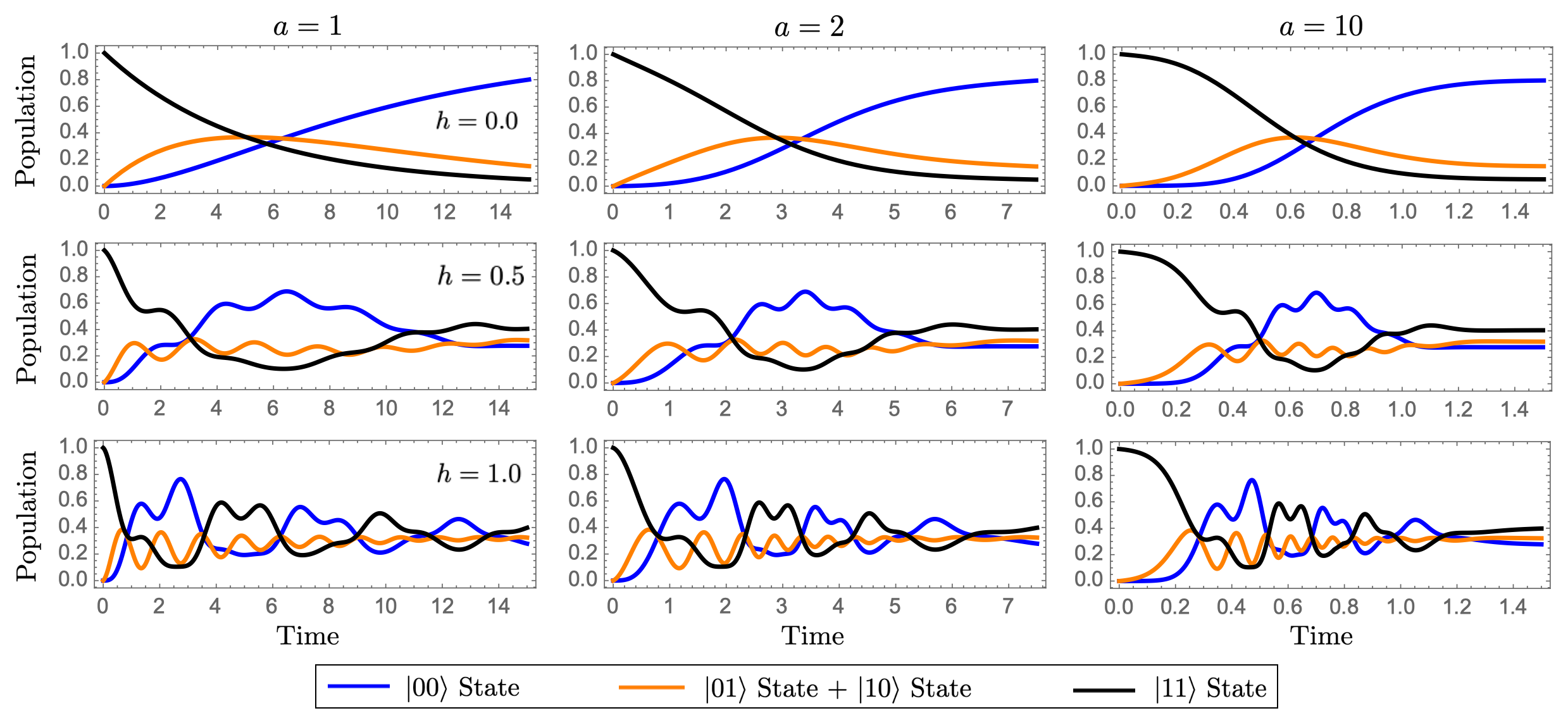}
\caption{Dissipative transverse field Ising model. Time evolution of the populations of the states $\ket{00}$ (blue), $\ket{11}$ (black) and the sum of the populations of the states $\ket{01}$ and $\ket{10}$ (orange). For the reference dynamics (first column), the transverse magnetic-field strengths $h$ are set as constants, with the nearest-neighbor coupling strength $J=1.0$ and the decay rate $\gamma = 0.1$. The second and third columns show the results of the corresponding TR dynamics for $a=2$ and $a=10$, respectively. In all cases, we set $\ket{11}$ as the initial state.}
\label{TFIM}
\end{center}
\end{figure*}

In the second and third columns of Fig.~\ref{amplitudedamping} we display the corresponding TR dynamics in which the time duration is shortened by a factor of 2 ($a=2$) and 10 ($a=10$), respectively. To generate these dynamics, the reference Liouvillian must be changed as in Eq.~(\ref{liouvillians}). Since the control parameters $\delta$, $\Omega$ and $\gamma$ are constant in the reference processes, the changes to be performed in order to generate the fast dynamics are simply $\delta \rightarrow \delta \dot{f}(t)$, $\Omega \rightarrow \Omega \dot{f}(t)$ and $\gamma \rightarrow \gamma \dot{f}(t)$, where for the $a=2$ and $a=10$ cases we have $\dot{f}(t) = 2-\cos(4 \pi t/t_f)$ and $\dot{f}(t) = 10 -9 \cos(20 \pi t/t_f)$, respectively. With no driving field ($\Omega = 0$), the excited state population simply decays monotonically to zero, as can be seen in the first line of Fig.~\ref{amplitudedamping}. In the nonzero driving case ($\Omega > 0$), oscillations of the populations are observed before an asymptotic stabilization. 

If we analyze the behavior of the processes without paying attention to the time duration, we note that the dynamics tend to develop more rapidly at intermediate times ($t \sim t_f/2a$) than at earlier ($t \sim 0$) and later times ($t \sim t_f/a$) for increasing values of $a$. This is because $f(t)$, which dictates the time evolution of the TR dynamics [see Eq.~(\ref{trroute})], grows more and more rapidly for intermediate times as $a$ increases (see Fig.~\ref{TRroutes}). It can also be seen that the trajectories of the reference and TR dynamics are identical, and that these dynamics are related to each other by a reparametrization of time, in agreement with the result of Eq.~(\ref{trroute}). The {\it Mathematica} notebook that generates the results displayed in Fig.~\ref{amplitudedamping} is available in the Supplemental Material \cite{sm}.

The experimental realization of the fast dynamics in the context of a two-level atomic system involves two parts: the unitary control, which is related to the realization of $\delta(t)=\delta \dot{f}(t)$ and $\Omega(t) = \Omega \dot{f}(t)$, and the nonunitary control, which corresponds to the manipulation of $\gamma(t) = \gamma \dot{f}(t)$. The constants $\delta$, $\Omega$ and $\gamma$ are the reference parameters, and $\dot{f}(t)$ is a smooth function of $t$. The unitary control can be realized by exposing the atoms to a laser beam whose intensity and frequency vary in time according to $\dot{f}(t)$. We call attention to the fact that the Rabi frequency $\Omega(t)$ and the detuning $\delta(t)$ scale proportionally to the intensity and frequency of the incident radiation field, respectively \cite{fox}. This type of unitary quantum control with atoms is commonly realized with current technology \cite{vitanov,saffman}. On the other hand, the implementation of the nonunitary control requires time-dependent manipulation of the coupling between the atomic system and its environment; a task generally more challenging than controlling Hamiltonian parameters \cite{celeri}. Nonetheless, engineered environments enabling such control have been experimentally demonstrated, for instance, in trapped-ion systems. In those experiments, a uniform oscillating electric field applied along the trap axis served as a tunable environment \cite{myatt}. Additionally, the use of squeezed vacuum fields as controlled environments for atomic systems has also been experimentally realized \cite{murch}.

We next apply our method to the transverse-field
Ising model (TFIM) with two sites, which is a four-level system. The Hamiltonian of the TFIM is
\begin{equation}
H = -J \sum_{k} \sigma^{(k)}_z \sigma^{(k+1)}_z - h \sum_{k} \sigma^{(k)}_x, 
\end{equation}
where $J$ is the nearest-neighbor coupling strength and $h$ is the transverse
magnetic-field strength.
Here, we assume the influence of the environment as described by the Lindblad damping operator $\sigma^{(k)}_{-}$ acting independently at each site $k$, with a decay rate $\gamma$. Fig.~\ref{TFIM} shows the population dynamics of the states $\ket{00}$ and $\ket{11}$ and the sum of the populations of the states $\ket{01}$ and $\ket{10}$ for different parameter settings, where we set the initial state to $\ket{11}$. Our calculations showed that the populations of the states $\ket{01}$ and $\ket{10}$ are the same in all processes. The reference ($a=1$) dynamics are displayed in the first column for some fixed values of $h$ to illustrate different dynamical behaviors. In these processes, we set $J=1$, $\gamma =0.1$, and $t_f=15$ in adimensional units. There are no population oscillations in the absence of the magnetic field ($h=0$). In contrast, population oscillations with increasing frequency take place as $h$ increases. Our simulations also revealed that the asymptotic values of the populations are independent of the field strength. The reference dynamics results agree with Refs.~\cite{kamakari,schli}.

The engineered TR dynamics for $a=2$ and $a=10$ are shown in the second and third columns of Fig.~\ref{TFIM}, respectively. These fast dynamics are generated from the reference ones by means of the application of the TR Liouvillian according to Eq.~(\ref{liouvillians}). Once the reference dynamics have fixed values for the control parameters, the required changes are $J \rightarrow J \dot{f}(t)$, $h \rightarrow h \dot{f}(t)$ and $\gamma \rightarrow \gamma \dot{f}(t)$. As we can see, our method successfully  provided open quantum dynamics, which are $a$ times faster than the corresponding reference processes, following the same trajectories. The property that TR processes develop more rapidly when $t \sim t_f/2a$ than when $t \sim 0$ and $t \sim  t_f/a$ is also evident in this example. In the Supplemental Material \cite{sm}, we have provided a {\it Mathematica} notebook that details the calculations of the results shown in Fig.~\ref{TFIM}.  

This application is also experimentally feasible. The unitary part requires time-dependent control of both the coupling strength, $J(t) = J \dot{f}(t)$, and the transverse magnetic field, $h(t) = h \dot{f}(t)$. Arrays of individually trapped Rydberg atoms are well-established as a versatile platform for simulating the quantum Ising model \cite{labuhn}. In such systems, the tunable distance between adjacent atoms effectively plays the role of the interaction strength, while the Rabi frequency of the laser coupling involving the ground and Rydberg states of the atoms serves as the transverse magnetic field. Both of these parameters have been demonstrated to be controllable with high precision using current experimental techniques \cite{bern}. The nonunitary control follows a similar strategy to that described in the previous example. In the following section, we present a reformulation of our method to accommodate the implementation of time-independent environments in the design of fast dynamics.

\section{Accelerated processes with time-independent environment}
\label{results2}

Although the structure of the Lindblad operators is time-independent, the rates $\gamma(t)$ in the TR processes introduced in the previous section are time-dependent. Physically, this implies that precise and rapid control over the environment degrees of freedom is required to generate the designed time control of the transition rates. As discussed in the previous section, this has been a challenging task. As such, it is of interest to explore methods for accelerating Lindblad dynamics without relying on time control of the environment properties. This is the focus of the present section. To achieve this, we employ a linear time-rescaling function, $g(t) = at$. This function satisfies properties (i) and (ii) outlined in Sec.~\ref{theory}, but not properties (iii) and (iv). As a result, the TR dynamics generated by $g(t)$ are indeed faster than the reference dynamics, but they involve different initial and final Liouvillian superoperators. This implies that the system will be subjected to different initial and final Hamiltonians, and that the initial and final environmental conditions will also differ. However, we will see that if the properties of the environment are time-independent in the reference dynamics, they will remain time-independent in the TR dynamics.

\begin{figure}[htb]
\includegraphics[width=0.47\textwidth]{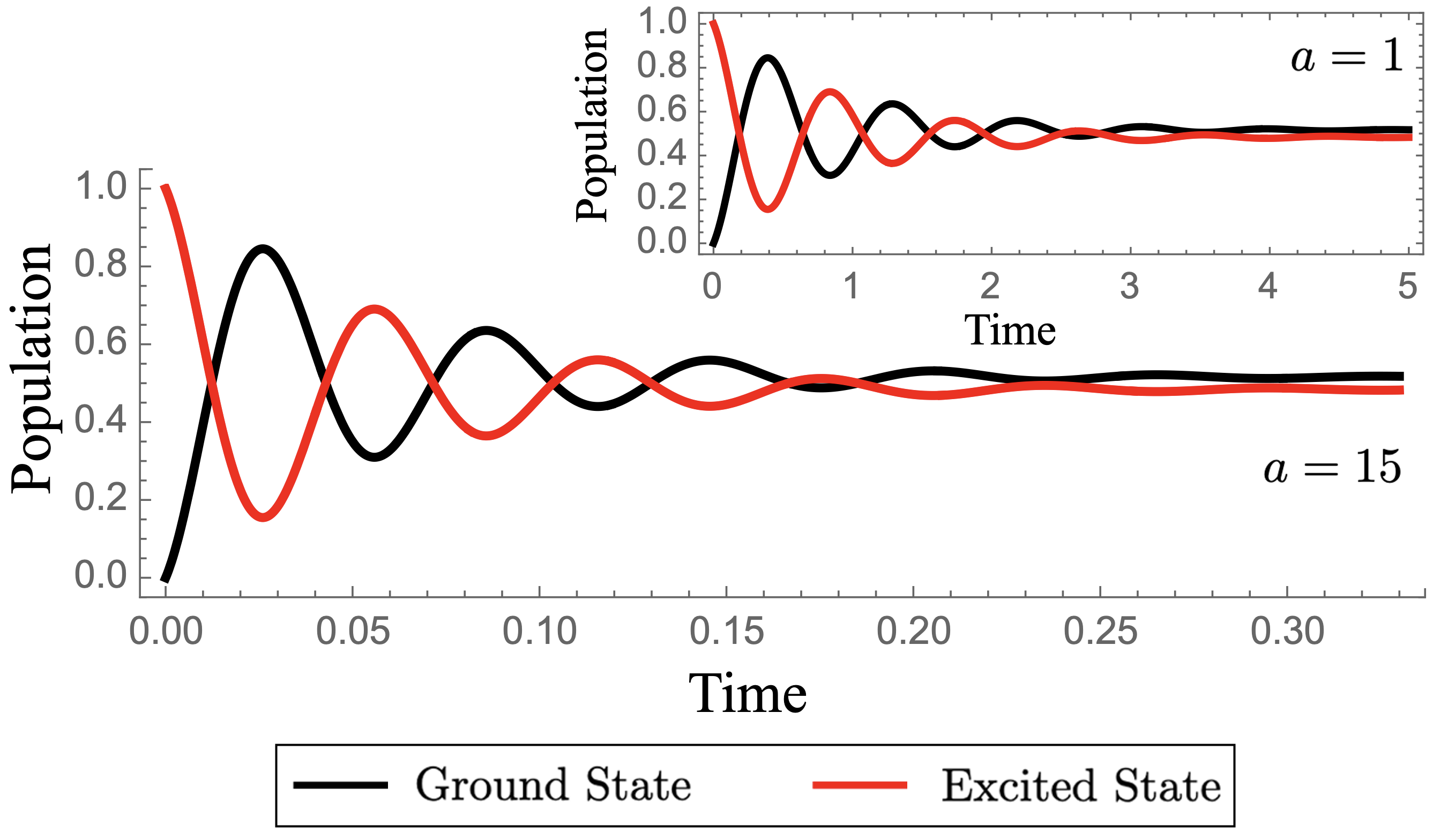}
\caption{Driven two-level system subjected to the amplitude damping channel. Population dynamics of the ground (black) and excited (red) states are shown. The main plot displays the behavior of the accelerated dynamics, which has a time duration 15 times ($a = 15$) shorter than the reference process ($a = 1$), shown in the inset. For the reference case, we use the parameters $\delta = 0.5$, $\Omega = 7$, and $\gamma = 1.5$. In the accelerated process, the corresponding parameters are $\delta = 7.5$, $\Omega = 105$, and $\gamma = 22.5$.}  
\label{TRroutes4}
\end{figure}

\begin{figure}[htb]
\includegraphics[width=0.47\textwidth]{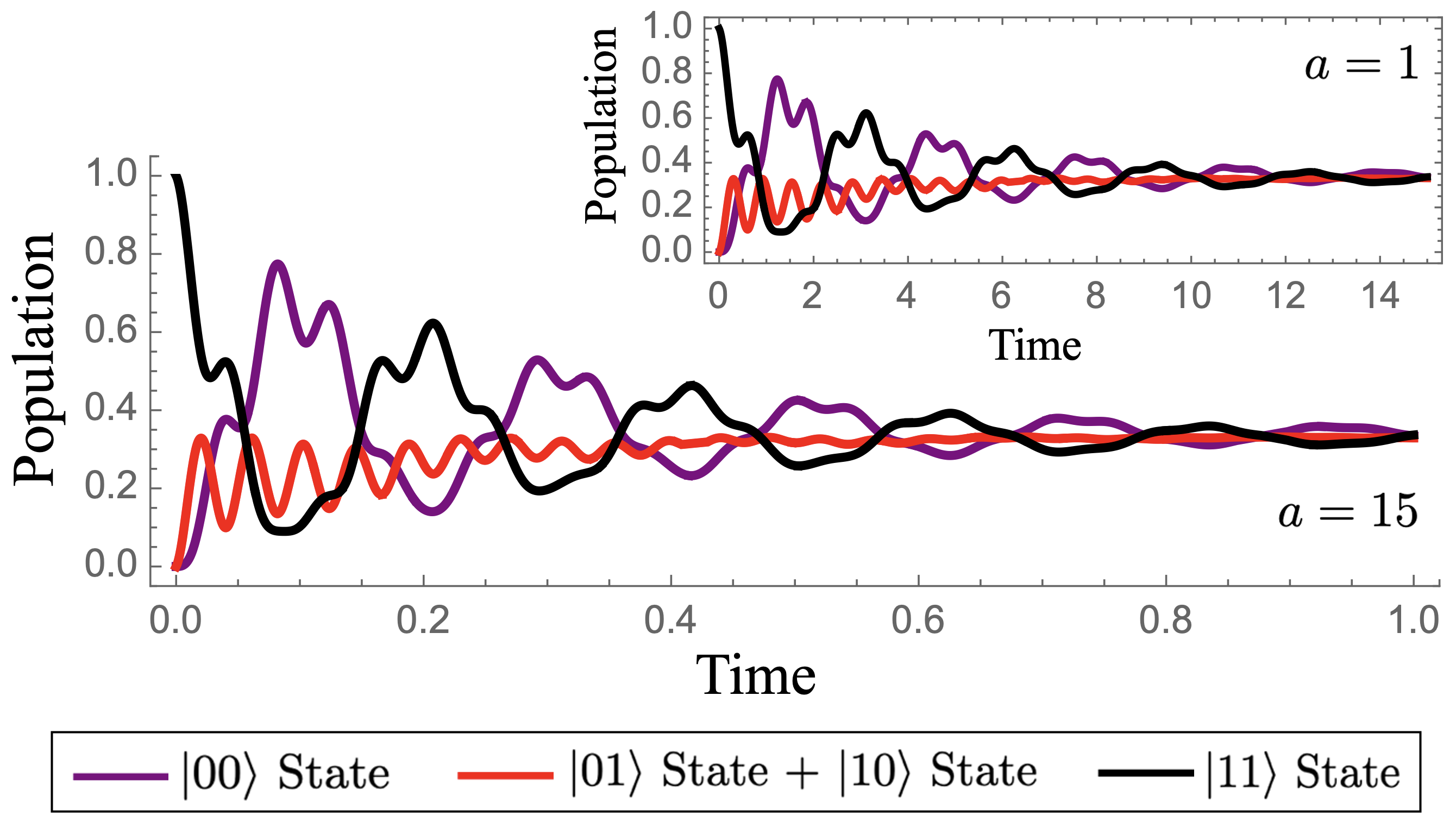}
\caption{Dissipative transverse field Ising model. Time evolution of the populations of the states $\ket{00}$ (purple), $\ket{11}$ (black), and the sum of the populations of the states $\ket{01}$ and $\ket{10}$ (red). The main plot shows a fast process accelerated by a factor of 15 ($a = 15$) compared to the reference process ($a = 1$), which is shown in the inset. The parameters used in the reference process are $J = 3$, $h = 2$, and $\gamma = 0.2$. The corresponding parameters for the accelerated process are $J = 45$, $h = 30$, and $\gamma = 3$.}
\label{TRroutes5}
\end{figure}

Following Eq.~(\ref{liouvillians}), the Liouvillian that generates the TR process now relates to the reference one as
\begin{align}
\label{liouvillians2}
\tilde{\mathcal{L}}(t) = \mathcal{L}[g(t)]\dot{g}(t)= a \mathcal{L}[g(t)].
\end{align}
This results in a transformation of the Hamiltonian given by $H(t) \rightarrow \mathcal{H}(t) = a H[g(t)]$, and similarly for the rates: $\gamma_k(t) \rightarrow a \gamma_k[g(t)]$. With this, if the rates are time-independent in the reference process, they remain time-independent in the accelerated dynamics, transforming as $\gamma_k \rightarrow a \gamma_k$. Therefore, to generate open dynamics that are $a$ times faster than the reference, one requires an environment with the density of particles increased by a factor of $a$, thereby enhancing the decay rates by the same factor. As previously mentioned, in the current framework, the initial and final Hamiltonians of the accelerated dynamics differ from those of the reference process. However, this is not a big problem, since the reference and TR Hamiltonians commute at the times when they generate the same state $\ket{\rho(t)}$, i.e., at times $t$ and $g^{-1}(t)$, respectively. In fact, we have
\begin{equation}
\label{commutator}
[H(t), \mathcal{H}[g^{-1}(t)]] = [H(t), aH(t)] = 0.    
\end{equation}
This commutation relation implies that the reference and TR Hamiltonians share the same set of eigenstates at the corresponding times. Therefore, if the state of the system is stationary at the initial and final times of the reference dynamics, the same will hold for the initial and final times of the accelerated dynamics.

This commutation relation also ensures that the dynamical properties of the state of the system throughout the entire reference process is preserved in the fast processes generated using the function $g(t)$. In general, two main differences arise when replacing the function $f(t)$ studied in the previous sections with $g(t)$. On the one hand, the TR processes designed via $g(t)$ require a different Hamiltonian at all times. On the other hand, there is no need for time-dependent control of the environment properties. In both cases the system state follows the same trajectory in Hilbert space as in the reference process at a rate which is $a$ times faster. In Figs.~\ref{TRroutes4} and~\ref{TRroutes5}, we illustrate the effect of the time-rescaling method using $g(t)$ for both the driven two-level system in an amplitude damping channel and the dissipative transverse field Ising model, in order to compare with results of the previous section. It is evident that the reference processes were perfectly accelerated in both cases. Furthermore, since the action
of the environment in these cases is time-independent, the designed fast processes also inherit this property, which significantly simplifies the practical implementation. 

\section{Quantum Speed Limit}
\label{qslimit}

The quantum speed limit (QSL) refers to the minimum time required for a quantum system to evolve from a given state to another under specific physical conditions (energy cost to realize the process and type of control) \cite{deffner}. One of the first QSL results for closed quantum systems was established by Mandelstam and Tamm in 1945 \cite{MT}, who showed that the evolution time is bounded from below by the inverse of the energy uncertainty of the system in the form $\Delta t \geq t_{QSL} = \pi \hbar / (2 \Delta H)$, where $\Delta H = \sqrt{\langle H^2 \rangle - \langle H \rangle^2}$. An alternative bound was later derived by Margolus and Levitin \cite{ML}, expressed as $\Delta t \geq t_{QSL} = \pi \hbar / (2 \langle H \rangle)$. Together, these two results define the fundamental lower bound on the evolution time between two orthogonal states in an isolated quantum system. For open quantum systems, however, the first results on quantum speed limits only emerged in 2013 \cite{campo, taddei, deffner2}.

In Ref.~\cite{campo}, del Campo {\it et al.} presented a generalization of the Mandelstam-Tamm bound for open quantum systems governed by Lindblad dynamics. For time-independent Liouvillians $\mathcal{L}$, the bound takes the form
\begin{equation} \label{qslbound} \Delta t_{\theta} \geq t_{QSL} = \frac{4 \theta^2  \mathrm{Tr}{[\rho(0)]^2}}{\pi^2 \sqrt{\mathrm{Tr}{[\mathcal{L}^{\dagger}\rho(0)]^2}}}, \end{equation} where the parameter $\theta$ characterizes the relative purity between the initial and final states as
\begin{equation} \theta = \arccos f(t_f) = \arccos \left( \frac{\mathrm{Tr}[\rho(0)\rho(t_f)]}{\mathrm{Tr}{[\rho(0)]^2}} \right). \end{equation} Here, $\rho(0)$ and $\rho(t_f)$ denote the initial and final states of the dynamics, respectively. The adjoint of the Liouvillian reads
\begin{equation}
\mathcal{L}^{\dagger} \rho = \frac{i}{\hbar} [H,\rho] + \sum_{k} \gamma_{k} \left( L^{\dagger}_{k} \rho L_{k} - \frac{1}{2} \{ L^{\dagger}_{k} L_{k}, \rho \}\right). 
\end{equation}

Let us now analyze how the QSL of a system governed by a Liouvillian $\mathcal{L}$ transforms under time-rescaling with the function $g(t) = at$, as discussed in the previous section. We assume that the reference dynamics is already the fastest possible evolution connecting $\rho(0)$ and $\rho(t_f)$; the so-called {\it quantum brachistochrone}. In this case, the minimum evolution time is given by the QSL time $t_{QSL}$ in Eq.~(\ref{qslbound}). After applying the TR transformation via $g(t)$, the total duration of the rescaled dynamics becomes $\Delta \tilde{t}_{\theta} = t_{QSL}/a$. This accelerated dynamics is generated by the Liouvillian $\tilde{\mathcal{L}}(t) = a \mathcal{L}$, as shown in Eq.~(\ref{liouvillians2}). Using Eq.~(\ref{qslbound}) to calculate the QSL time for the transformed Liouvillian, we obtain
\begin{equation} \label{trqslbound} \tilde{t}_{QSL} = \frac{4 \theta^2  \mathrm{Tr}{[\rho(0)]^2}}{\pi^2 \sqrt{\mathrm{Tr}{[\tilde{\mathcal{L}}^{\dagger}\rho(0)]^2}}} = \frac{4 \theta^2  \mathrm{Tr}{[\rho(0)]^2}}{a\pi^2 \sqrt{\mathrm{Tr}{[\mathcal{L}^{\dagger}\rho(0)]^2}}}, \end{equation} which leads to the result $\Delta \tilde{t}_{\theta} = t_{QSL}/a = \tilde{t}_{QSL}$.

In other words, if the original Liouvillian $\mathcal{L}*$ generates the fastest evolution between two states $\rho(0)$ and $\rho(t_f)$ under the constraint $\sqrt{\mathrm{Tr}{[\mathcal{L}^{\dagger}\rho(0)]^2}} = \nu$, then the rescaled Liouvillian $\tilde{\mathcal{L}}_* = a \mathcal{L}_*$ yields the fastest evolution within the class of dynamics characterized by $\sqrt{\mathrm{Tr}{[\mathcal{L}^{\dagger}\rho(0)]^2}} = a \nu$. More generally, by classifying dynamical regimes according to the quantity $\sqrt{\mathrm{Tr}{[\mathcal{L}^{\dagger}\rho(0)]^2}}$, we find that the quantum brachistochrone of one dynamical regime gives rise, via time rescaling, to the brachistochrones of other dynamical scenarios. This highlights the universality and versatility of the time-rescaling method.
\\\\

\section{Conclusion}
\label{conclusions}

We have formulated a universal framework to accelerate Lindblad dynamics, which generalizes the time-rescaling method \cite{bernardo} to open quantum systems. The method is simple, analytical, and has significant advantages over previous STA schemes. First, our approach provides exact fast processes that are driven through the same trajectory in Hilbert space as the reference protocol. In fact, the reference and the fast dynamics were shown to be related by a simple reparametrization of time [see Eq.~(\ref{trroute})]. Second, the engineered dynamics are Markovian and do not necessarily require time control of the environment properties. Third, calculations of the protocols do not require information about the (complex) eigenvalues of the reference Liouvillian, and the experimental realization does not involve extra control fields.

Another important feature of our approach is that the fast dynamics are generated by a Liouvillian that satisfies locality constraints, as long as this is also true for the reference process. This result is expected to be relevant to the study of dissipative quantum state engineering in many-body systems and the design of quantum many-body engines \cite{verstraete,koch}. We verified the efficacy of the method in two scenarios: a driven two-level system in a heat bath at zero temperature and the dissipative transverse field Ising model. Moreover, for a reference process that is known to take the minimal time evolution between two states, under a determined dynamical constraint, it was shown that the proposed method enables to find a family of optimal processes that comply with different values of the constraint.

{\it Acknowledgement.}-- The author acknowledges support from the Brazilian agencies Coordenação de Aperfeiçoamento de Pessoal de Nível Superior (CAPES) and Conselho Nacional de Desenvolvimento Cient{\'i}fico e Tecnol{\'o}gico (CNPq - Grant No. 307876/2022-5).   
\\

\end{document}